\documentclass[aps,prd,11pt,superscriptaddress,nofootinbib]{revtex4-2}
\usepackage[utf8]{inputenc}
\usepackage{amsmath}
\usepackage{amssymb}
\usepackage{bm}
\usepackage{graphicx}
\usepackage{subfigure}
\usepackage{hyperref}
\usepackage{comment}
\usepackage{ulem}
\usepackage{CJK}
\hypersetup{
  colorlinks=true,
  linkcolor=red,
  citecolor=blue,
}
\begin{document}
\begin{CJK*}{UTF8}{gbsn}
\title{Gauge transformation of scalar induced tensor perturbation during matter domination}
\author{Arshad Ali}
\email{aa$_$math@hust.edu.cn}
\affiliation{School of Physics, Huazhong University of Science and Technology,
Wuhan, Hubei 430074, China}
\author{Yungui Gong(龚云贵)}
\email{Corresponding author. yggong@hust.edu.cn}
\affiliation{School of Physics, Huazhong University of Science and Technology,
Wuhan, Hubei 430074, China}
\author{Yizhou Lu(卢一洲)}
\email{louischou@hust.edu.cn}
\affiliation{School of Physics, Huazhong University of Science and Technology,
Wuhan, Hubei 430074, China}

\begin{abstract}
The scalar-induced secondary gravitational wave as  the stochastic gravitational background is a useful tool to study the physics in the early universe. 
We study the scalar-induced tensor perturbations at second-order during matter domination in seven different gauges.
We obtain the results in six other gauges from that in the Newtonian gauge 
using the gauge transformation law of the scalar-induced tensor perturbation.
We find that the kernel functions $I_{\chi}$ in the synchronous and comoving orthogonal gauges are the same if the residual gauge modes in these two gauges 
are eliminated.
By identifying the oscillating terms $\sin x$ and $\cos x$ in the scalar-induced tensor perturbations as the scalar-induced secondary gravitational waves, 
we find that its energy density is actually gauge independent. 
The energy density $\rho_{\text{GW}}\propto a^{-4}$, or $\Omega_{\text{GW}}\propto a^{-1}$
in the matter-dominated era, and the scalar-induced secondary gravitational waves behave as radiation.
\end{abstract}


\maketitle
\end{CJK*}

\section{Introduction}
The discovery of gravitational waves (GWs)
 from mergers of black holes (BHs) and neutron stars (NS) by the Laser Interferometer Gravitational-Wave Observatory (LIGO) Scientific Collaboration and  Virgo Collaboration \cite{TheLIGOScientific:2016agk,Abbott:2016blz,Abbott:2016nmj, Abbott:2017vtc,Abbott:2017oio,TheLIGOScientific:2017qsa, Abbott:2017gyy,LIGOScientific:2018mvr,Abbott:2020uma,LIGOScientific:2020stg,Abbott:2020khf,Abbott:2020tfl,Abbott:2020niy}
has marked the beginning of the era of astronomy of GWs. There are also GWs with cosmological origins, 
such as the primordial GWs generated during inflation,
the second-order GWs 
induced by primordial scalar perturbation, as well as
GWs generated from a cosmic phase transition \cite{ Kosowsky:1992vn,Ananda:2006af,Baumann:2007zm,Saito:2008jc,Assadullahi:2009nf,Sathyaprakash:2009xs,Saito:2009jt,Bugaev:2009zh,Jedamzik:2010hq,Alabidi:2012ex,Alabidi:2013lya,Gao:2014pca,Gong:2014cqa,Guzzetti:2016mkm,Gong:2017qlj,Espinosa:2018eve,Cai:2018tuh,Cai:2018dig,Kohri:2018awv,Inomata:2018epa,Kamenshchik:2018sig,Cai:2019jah,DeLuca:2019ufz,Domenech:2019quo,Yuan:2019fwv,Yuan:2019udt,Yuan:2019wwo,Fu:2019vqc,Hajkarim:2019nbx,Tomikawa:2019tvi,Lu:2019sti,Lu:2020diy,Nakamura:2019zbe, Inomata:2019yww,Inomata:2019zqy,Gong:2019mui,Dalianis:2020cla,Domenech:2020kqm,Inomata:2020lmk,Zhou:2020ils,Lin:2020goi,Zhang:2020ptw,Pi:2020otn,Yi:2020kmq,Yi:2020cut,Braglia:2020eai,Ragavendra:2020sop,Gundhi:2020zvb}.
The power spectrum of primordial curvature perturbations at large scales is nearly scale-invariant with the amplitude $A_s=2.1\times 10^{-9}$ at the pivot scale $k=0.05~\mathrm{Mpc}^{-1}$ \cite{Akrami:2018odb}, while the small-scale ones remain to be explored. 
If the primordial scalar perturbations at small scales
are large enough (the amplitude of the power spectrum needs to be at least $0.01$), 
then a sizable amount of secondary GWs will be induced during the radiation domination (RD) and the matter domination (MD) due to the mixing of tensor and scalar perturbations. 
The scalar induced GWs (SIGWs) at second-order contribute to the stochastic gravitational-wave background,
so it is possible to extract the information about small scale primordial scalar perturbations from the detection of SIGWs.
In other words, SIGWs can be used to probe the thermal history of the universe and to understand the physics during inflation \cite{Boyle:2005se,Hajkarim:2019nbx,Domenech:2020kqm}.

In contrast to tensor perturbations at the first order,
the second-order tensor perturbations are gauge-dependent, so SIGWs may depend on the gauge choice \cite{Arroja:2009sh,Hwang:2017oxa,Tomikawa:2019tvi,Inomata:2019yww,Yuan:2019fwv,DeLuca:2019ufz,Lu:2020diy,Chang:2020iji,Chang:2020mky},
even though many choices of the gauge-invariant tensor perturbation at second order can be constructed in a specific gauge \cite{Bruni:1996im,Mollerach:2003nq,Bartolo:2004ty,Nakamura:2004rm,Nakamura:2006rk,Malik:2008im,Arroja:2009sh,Domenech:2017ems,Blaut:2019fxb,DeLuca:2019ufz,Chang:2020tji,Domenech:2020xin}.
This means that we need to calculate the second-order tensor perturbations in each gauge. 
However, the production of SIGWs was usually discussed in the Newtonian gauge.
It is necessary to discuss SIGWs in other gauges.
During RD, the energy densities of SIGWs in the Newtonian, synchronous, and 
uniform curvature gauge were found to be the same \cite{Yuan:2019fwv,DeLuca:2019ufz,Inomata:2019yww}.
The energy density of SIGWs in the synchronous gauge during both RD and MD was discussed in \cite{Inomata:2019yww}.
For a general background with
the constant equation-of-state $w$,
particularly, for RD with $w=1/3$, and MD with $w=0$, the SIGWs were calculated in the Newtonian, comoving, and uniform curvature gauges \cite{Tomikawa:2019tvi}.
In the previous paper \cite{Lu:2020diy}, we derived a general formula for the calculation of SIGWs during RD in an arbitrary gauge and obtained the results for SIGWs in the uniform
curvature gauge, the synchronous gauge, the comoving gauge, the comoving orthogonal gauge,
the uniform density gauge, and the uniform expansion gauge from the result in the Newtonian gauge using the coordinate transformation. 
As expected, the energy density of SIGWs in different gauge is not invariant in general.
But a physical observable like SIGWs should not depend on the gauge choice.

To resolve the problem of the gauge-dependence of SIGWs, we recall the property of GWs
to distinguish SIGWs from the second-order tensor perturbation. 
In general relativity, GWs propagate at the speed of light, and the freely propagating GWs oscillate like $\sin(k\eta)$ or $\cos(k\eta)$ and their energy densities decay as $a^{-4}$,
so only the terms with oscillating behaviors like $\sin(k\eta)$ or $\cos(k\eta)$ in the scalar-induced tensor perturbations should be identified as SIGWs. 
The terms behave other than $\sin(k\eta)$ and $\cos(k\eta)$ are not SIGWs, and they arise because of the mixing of tensor and scalar perturbations.
This point of view was adopted in \cite{Inomata:2019yww} to show SIGWs in the Newtonian and synchronous gauges are the same. 
Similarly, by arguing that the source of SIGWs is not active for modes deep inside the horizon, 
it was shown that the energy density for SIGWs at small scales is well behaved and invariant under a set of reasonable gauge transformation in general cosmological backgrounds except MD \cite{Domenech:2020xin}. 
If we consider the contribution from terms oscillating as $\sin(k\eta)$ and $\cos(k\eta)$ only, 
then, using the results obtained in Ref. \cite{Lu:2020diy}, it is easy to show that SIGWs during RD are the same in
the uniform curvature gauge, the synchronous gauge, 
the comoving gauge, the comoving orthogonal gauge,
the uniform density gauge, the uniform expansion gauge, and the Newtonian gauge. 
In a general cosmological background with a constant equation-of-state $w\neq 0$, 
the Newtonian potential oscillates with the oscillation frequency $\sqrt{w}k$ and decays as $\eta^{-3(1+w)/(1+3w)}$ inside the horizon in the Newtonian gauge. 
However, the MD with $w=0$ is a special case. 
During MD, the Newtonian potential is a constant in all scales and does not decouple from the tensor perturbation at the second order.
Therefore, SIGWs in MD deserve to be further investigated.

In this paper, we discuss the energy density of SIGWs during MD in seven various gauges and show that SIGWs with the oscillating behaviors  $\sin(k\eta)$ and $\cos(k\eta)$ are the same in these gauges.
The paper is organized as follows. 
The basic formulas used to 
calculate SIGWs and discuss the gauge
transformation are given in Sec. \ref{sec.2}. 
We also provide the prescription to obtain the
expressions in other gauges from the Newtonian gauge result by using  the gauge transformation
of the second-order tensor perturbation in MD. 
In Sec. \ref{sec.3}, we derive the kernels $I(u,v,x)$ analytically in different gauges. Then we calculate the energy density of SIGWs with the behaviors $\sin(k\eta)$ or $\cos(k\eta)$ in the seven gauges and discuss the gauge independence of the result.
The summary of our results is presented 
in Sec. \ref{sec4}.

\section{Formulation of SIGWs and gauge transformations}
\label{sec.2}
In this section, we discuss the formalism of the second-order SIGWs. We begin with the following perturbed metric
	\begin{equation}
	\label{meric}
	\begin{split}
\mathrm{d}s^{2}=&-a^2 (1+2 \phi) \mathrm{d} \eta^{2}+2a^2B_{,i} \mathrm{d} x^{i} \mathrm{d} \eta\\
&+a^2\left[(1-2\psi)\delta_{i j}+2E_{,ij}+\frac12 h_{ij}^{\mathrm{TT}}\right] \mathrm{d} x^{i} \mathrm{d} x^{j},
 \end{split}
\end{equation}
where the metric perturbations include the first-order scalar perturbations $\phi$, $\psi$, $B$, and $E$,
and the second-order tensor perturbation $h_{ij}^{\mathrm{TT}}$,
which is transverse and traceless: $h^{\mathrm{TT}}_{ii}=\partial_i h^{\mathrm{TT}}_{ij}=0$.
The first-order tensor perturbation and the vector perturbations are not taken into account because we discuss SIGWs only.
In the forthcoming derivations of SIGWs, we assume that the production of induced GWs begins long before the horizon reentry.

\subsection{The generation of SIGWs}
The equation of motion for the transverse traceless tensor mode $h^{\mathrm{TT}}_{ij}$ at the second-order can be derived straightforwardly from the perturbed Einstein's equation $G_{\mu\nu}=8\pi GT_{\mu \nu}$  as
\begin{align}
\label{theeq}
  h_{ij}^{\mathrm{TT}\prime\prime}+2\mathcal{H}h_{ij}^{\mathrm{TT}\prime}-\nabla^2h_{ij}^{\mathrm{TT}}=4\mathcal{T}_{ij}^{lm}s_{lm},
\end{align}
where the prime stands for the derivative with respect to the conformal time  $\eta$, $\mathcal{H}=a^\prime(\eta)/a(\eta)$ is the comoving Hubble parameter, and $s_{lm}$ is the source term given below in Eq. \eqref{source}. The projection tensor $\mathcal{T}_{ij}^{lm}$ acting on the source term extracts the transverse and traceless part and will be discussed below. In this paper,
we consider the production of SIGWs in MD only, where $a= \eta^2$ and $\mathcal{H}= 2/\eta $.

The Fourier components of the transverse and traceless parts of the tensor perturbations in terms of the polarization tensors are defined as 
 \begin{equation}
 \label{hijkeq1}
      h_{ij}^{\mathrm{TT}}(\bm y,\eta)=\int\frac{\mathrm{d}^3k}{(2\pi)^{3/2}}e^{i{\bm k}\cdot {\bm y}}\Bigl[h^+_{\bm k}(\eta)\mathbf e^+_{ij}+{h}^\times_{\bm k}(\eta) \mathbf e^\times_{ij}\Bigr],
 \end{equation}
 where $\bm y$ are spatial coordinates.
In terms of the orthonormal bases ${\mathbf e}$ and $\bar{\mathbf e}$ orthogonal to the wave vector $\bm{k}$,
with ${\bm k}\cdot {\mathbf e}={\bm k}\cdot \bar{\mathbf e}={\mathbf e}\cdot \bar{\mathbf e}=0$ and $|{\mathbf e}|=|\bar{\mathbf e}|=1$,
 the plus and cross polarization bases are defined as
  \begin{equation}
 \label{poltensor1}
 \begin{split}
      \mathbf e^+_{ij}=\,\,&\frac{1}{\sqrt{2}}\left[\mathbf e_i \mathbf e_j-\bar{\mathbf e}_i \bar{\mathbf e}_j\right],\\
      \mathbf e_{ij}^\times=\,\,&\frac{1}{\sqrt{2}}\left[\mathbf e_i\bar{\mathbf e}_j+\bar{\mathbf e}_i \mathbf e_j\right].
 \end{split}
 \end{equation}
 The polarization tensors \eqref{poltensor1} are transverse and traceless as $k_i \mathbf e^+_{ij}=k_i \mathbf e^\times_{ij}=0$ and
$\mathbf e^+_{ii}=\mathbf e^\times_{ii}=0$.
In the Fourier space, the projection tensor is expressed as
 \begin{equation}
 \label{projeq}
     \mathcal{T}_{ij}^{lm}=[\mathbf e_{ij}^+ \mathbf e^{+lm}+\mathbf e_{ij}^\times \mathbf e^{\times lm}],
 \end{equation}
and the solution to Eq. \eqref{theeq}  for either polarization $e_{ij}^t$ reads

\begin{equation}
\label{hsolution1}
  h^t_{\bm k}(\eta)=4 \int_0^x\mathrm{d}\tilde{x}\frac{a(\tilde{\eta})}{a(\eta)}\frac{1}{k}G_{k}(\eta,\tilde{\eta}) S_{\bm k}^{t},
\end{equation}
where the Green's function $G_{k}(\tilde{\eta},\eta)$ to Eq. \eqref{theeq}
in MD is
\begin{equation}\label{greenfunction}
G_k(\tilde{\eta},\eta)=\frac{(1+x\tilde{x})\sin(x-\tilde{x})-(x-\tilde{x})\cos(x-\tilde{x})}{k x \tilde{x}},
\end{equation}
$\tilde x=k\tilde{\eta}$, $x=k\eta$, 
the source  $S_{\bm k}^{t}=\mathbf{e}^t_{ij}s^{ij}(\bm k,\eta)$ for either polarization $t=+$ or $\times$ is
 \begin{equation}
\label{sfrel1}
S_{\bm k}^{t}(\eta)=\int\frac{\mathrm{d}^3 p}{(2\pi)^{3/2}}\zeta(\bm p)\zeta(\bm k-\bm p)\mathbf{e}^t_{ij}p^i p^j f(u,v,x),
\end{equation}
$\zeta(\bm k)$ is the primordial curvature perturbation, $u=p/k$, $v=|\bm k-\bm{p}|/k$ and $f(u,v,x)$ will be given below in the next subsection. 
For convenience we introduce the integral kernel \cite{Ananda:2006af,Baumann:2007zm}
\begin{equation}
\label{I_int}
    I(u,v,x)=\int_0^x\mathrm{d}\tilde{x}\frac{a(\tilde{\eta})}{a(\eta)}kG_{k}(\eta,\tilde{\eta})f(u,v,\tilde x),
\end{equation}
to express the solution $h^t_{\bm k}(\eta)$,
\begin{equation}
\label{hsolution}
  h^t_{\bm k}(\eta)=4 \int\frac{\mathrm{d}^3p}{(2\pi)^{3/2}} \mathbf e^t_{ij}p^i p^j\zeta(\bm p)\zeta(\bm k-\bm p)\frac{1}{k^2}I(u,v,x).
\end{equation}

For free propagating GWs without the source, we get
the decaying oscillating solution
\begin{equation}
\label{hksol1}
  h^t_{\bm k}(\eta)=\frac{3(\sin x-x\cos x)}{x^3},
\end{equation}
with the initial condition $h^t_{\bm k}(0)=1$.

\subsection{The source term for SIGWs}
The second-order source term $s_{ij}$ for SIGWs in Eq. \eqref{theeq} that comes from the first-order scalar perturbation is \cite{Lu:2020diy}
\begin{widetext}
\begin{equation}
\label{source}
\begin{split}
 s_{ij}=\,\,&-\psi_{,i}\psi_{,j}-\phi_{,i}\phi_{,j}
 +\sigma_{,ij}\left(\phi^\prime+\psi^\prime-\nabla^2\sigma\right)
 -\left(\psi_{,i}^\prime\sigma_{,j}+\psi_{,j}^\prime\sigma_{,i}\right)
 +\sigma_{,ik}\sigma_{,jk}-2\psi_{,ij} \left(\phi+\psi\right)\\
 &-\frac{2}{\mathcal{H}'-\mathcal{H}^2}(\psi'+\mathcal{H}\phi)_{,i} (\psi'+\mathcal{H}\phi)_{,j}
 +2\psi_{,ij}\nabla^2E
 -2E_{,ij}\left(\psi^{\prime\prime}+2\mathcal{H}\psi^\prime-\nabla^2\psi\right)\\
 &-2\left(\psi_{,jk}E_{,ik}+\psi_{,ik}E_{,jk}\right)
 +2\mathcal{H}(\psi_{,i}E_{,j}^\prime+\psi_{,j}E_{,i}^\prime)+\psi_{,i}^\prime E_{,j}^\prime+\psi_{,j}^\prime E_{,i}^\prime
  +\psi_{,i}E_{,j}^{\prime\prime}+\psi_{,j}E_{,i}^{\prime\prime}\\
 & +E_{,ik}^\prime E_{,jk}^\prime
 -E_{,ikl}E_{,jkl}-2E_{,ij}^\prime\psi^\prime
 -E_{,ijk} \left(E^{\prime\prime}+2\mathcal{H}E^\prime-
 \nabla^2E\right)_{,k},
 \end{split}
 \end{equation}
 \end{widetext}
where the anisotropic stress tensor $\Pi_{ij}$ of the matter fluid
is assumed to be zero, the terms proportional to $\delta_{ij}$ are omitted because they do not contribute to the transverse and traceless part,
$\sigma= E^\prime-B$ is the shear potential, and
$\rho_0$ and $P_0$ are the background values
of energy density and pressure for the matter fluid.
The detailed discussion of these variables is presented in
Appendix \ref{app.A}.
In gauges with $E=0$, the above equation \eqref{source}
reduces to that given in \cite{Gong:2019mui,DeLuca:2019ufz,Hwang:2017oxa} with vanishing anisotropic stress.  In general,
we need to use Eq. \eqref{source} instead. Particularly, we should include all the terms involving $E$ in the synchronous gauge.

Come back to the function $f(u,v,x)$ in Eq. \eqref{sfrel1} which contains the source information and is gauge-dependent.
The explicit expression of the source function $f(u,v,x)$ is 
\begin{equation}
    \label{tilde_f1}
    f(u,v,x)=\,\,\frac12\times\frac{9}{25}\Bigl(\Tilde{f}(u,v,x)+\tilde{f}(v,u,x)\Bigr),
\end{equation}
where
\begin{widetext}
	\begin{equation}
\label{tilde_f}
	\begin{split}
\Tilde{f}(u,v,x)=\,\,&T_\psi(ux)T_\psi(vx)-T_\phi(ux)T_\phi(vx)-\frac{v}{u}T_\sigma(ux)\left[T_\phi^*(vx)+T_\psi^*(vx)+T_\sigma(vx)\right]-2\frac{u}{v}T_\psi^*(ux)T_\sigma(vx)
\\
&-\frac{1-u^2-v^2}{2uv}T_\sigma(ux)T_\sigma(vx)+2T_\psi(ux)T_\phi(vx)+ 2T_\psi(ux)T_E(vx)-\frac{1-u^2-v^2}{2uv}T_E^*(ux)T_E^*(vx)
\\
&+2\frac{u^2}{v^2}T_E(vx)\left[T_\psi^{**}(ux)+\frac{2\mathcal{H}}{ku}T_\psi^*(ux)+T_\psi(ux)\right]\\
&+\frac{2}{\mathcal{H}^2-\mathcal{H}^\prime}\left[kuT_\psi^*(ux)+\mathcal{H}T_\phi(ux)\right]\left[kvT_\psi^*(vx)+\mathcal{H}T_\phi(vx)\right]
\\
&+4\frac{\mathcal{H}}{kv}T_\psi(ux)T_E^*(vx)
+4\frac{u}{v}T_\psi^*(ux)T_E^*(vx)+2T_\psi(ux)T_E^{**}(vx) +2\left(\frac{1-u^2-v^2}{v^2}\right)T_\psi(ux)T_E(vx)\\
&-\left(\frac{1-u^2-v^2}{2uv}\right)^2T_E(ux)T_E(vx)
-\frac{1-u^2-v^2}{2u^2}T_E(ux)\left[T_E^{**}(vx)+2\frac{\mathcal{H}}{kv}T_E^*(vx)+T_E(vx)\right],
\end{split}
\end{equation}
\end{widetext}
where $T^*(x)=\mathrm{d}T(x)/\mathrm{d}x$ and the transfer functions $T(x)$ which 
relate the scalar perturbations to its primordial curvature perturbation
are defined as  \cite{Inomata:2019yww}
 \begin{gather}
 \label{defs}
   \sigma(\bm k,x)=\frac{3}{5}\zeta(\bm k)\frac{1}{k}T_\sigma(x),\\
   E(\bm k,x)=\frac{3}{5}\zeta(\bm k)\frac{1}{k^2}T_E(x),\\
   B(\bm k,x)=\frac{3}{5}\zeta(\bm k)\frac{1}{k}T_B(x),\\
   \psi(\bm k,x)=\frac{3}{5}\zeta(\bm k)T_\psi(x),\\
   \phi(\bm k,x)=\frac{3}{5}\zeta(\bm k)T_\phi(x).
 \end{gather}
From Eq. \eqref{tilde_f1}, it is obvious that the source function $f(u,v,x)$
is symmetric {\color{red}in} $u$ and $v$.

\subsection{The power spectrum of SIGWs}

The primordial curvature perturbation induces GWs in the MD era,
and the energy density of SIGWs is
\begin{equation}
\label{rhogw1}
    \rho_{\mathrm{GW}}=\frac{1}{32\pi G}\frac{1}{4a^2}\langle h'_{ij}h'_{ij}\rangle.
\end{equation}
We see that the contribution to the energy density from a constant $h_{ij}$ is zero due to the time derivative.
By using the tensor power spectrum $\mathcal{P}_h$ defined as
\begin{equation}
\label{hk}
\left\langle h_{\bm k_1}^{t_1}(\eta)h_{\bm k_2}^{t_2}(\eta)\right\rangle
=\frac{2\pi^2}{k_1^3}\delta_{t_1 t_2}\delta^3(\bm k_1+\bm k_2)\mathcal{P}_h(k_1,\eta),
\end{equation}
we get the energy density parameter $ \Omega_{\mathrm{GW}}(k,x)$ of SIGWs as
\begin{equation}
\label{omegagw}
    \Omega_{\mathrm{GW}}=\frac{\mathrm{d}\rho_{\mathrm{GW}}}{\rho_c\mathrm{d}\ln k}
    =\frac{1}{24}\left(
    \frac{x}{2}
    \right)^2\overline{\mathcal{P}_h(k,x)},
\end{equation}
where $t_i=+,\times$, an overbar stands for oscillatory average and $\rho_c=3H^2/8\pi G$ is
the critical energy density of the universe.
In deriving the second equality in Eq. \eqref{omegagw},
we use the fact that either polarization contributes equally
to the energy density and GWs are null waves, so we make the replacement
$|h_{\bm k}'(\eta)|^2=k^2|h_{\bm k}(\eta)|$ in the subhorizon limit with $k\gg \mathcal{H}$ \cite{Boyle:2005se}.

Combining Eqs. \eqref{hsolution} and \eqref{hk}, we get \cite{Kohri:2018awv,Espinosa:2018eve}
\begin{equation}
\label{PStensor}
\begin{split}
\mathcal{P}_h(k,x)=&4\int_{0}^\infty\mathrm{d}u\int_{|1-u|}^{1+u}\text{d}v I^{2}(u,v,x)\mathcal{P}_\zeta(uk)\\
&\qquad \times \mathcal{P}_\zeta(vk) \left[\frac{4u^2-(1+u^2-v^2)}{4uv}\right]^2,
 \end{split}
\end{equation}
where $\mathcal{P}_\zeta$ is the primordial scalar power spectrum.

\subsection{Newtonian gauge}
\label{newton}

To calculate the energy density explicitly, 
we need to choose a gauge.
The  SIGWs in Newtonian (Poisson) gauge during MD were studied in \cite{Baumann:2007zm,Assadullahi:2009nf,Hwang:2017oxa,Kohri:2018awv,Tomikawa:2019tvi,Inomata:2019yww}, we review the result in this subsection.
We introduce the transfer function $T$ to separate the time evolution by  defining $\phi(\bm k,\eta)=\phi(\bm k,0)T(\eta)$. 
In the Newtonian gauge where $B=E=0$, and ignoring anisotropic stress, we have $\phi_\mathrm{N}=\psi_\mathrm{N}=\Phi=\Psi$, in which
the Bardeen's potentials $\Phi$ and $\Psi$ defined in \eqref{bardeenphi} and \eqref{bardeenpsi} are given by $\Phi=\Psi=3\zeta/5$ on superhorizon scales.
Therefore,  $\phi_\text{N}(\bm k,0)=3\zeta(\bm k)/5$
and $T_{\text N}(x)=1$. The subscript ``$\mathrm{N}$" indicates that these quantities are evaluated in the Newtonian gauge.

In the Newtonian gauge, the source function in terms of the transfer functions $T_{\text N}$ for the gravitational potential $\phi_{\text N}$ reads 
\begin{equation}
\label{Nsource}
\begin{split}
f_{\text{N}}(v, u, x) = &\frac{6}{5}T_\mathrm{N}(v{x})T_\mathrm{N}(u{x})
+\frac{3}{25}  uv{x}^2 {T_\mathrm{N}^* (v{x})}{T_\mathrm{N}^* (u{x})}\\
& +\frac{6}{25} \left[v x T_\text{N}^*(vx) T_\text{N}(ux)+ux{T_\mathrm{N}^* (u{x})}T_\mathrm{N}(v{x}) \right].\\
\end{split}
\end{equation}
Using $T_{\mathrm{N}}(x)=1$,
we obtain
\begin{equation}\label{fNewton}
    f_{\mathrm{N}}(u,v,x)=\frac{6}{5}.
\end{equation}
Combining Eqs. \eqref{I_int}, \eqref{greenfunction} and \eqref{fNewton}, 
we get the explicit expression for the kernel $I_{\mathrm{N}}(u,v,x)$ \cite{Kohri:2018awv}
\begin{equation}\label{I_N}
   I_{\mathrm{N}}(u,v,x)=\frac{6}{5}+\frac{18(x\cos x -\sin x)}{5x^3}.
\end{equation}
Since $I_{\mathrm{N}}(u,v,x\rightarrow\infty)=6/5$,  Eq. \eqref{PStensor} tells us that at late times
$\mathcal{P}_h$ is a constant 
and the energy density of SIGWs is proportional to  $x^2$. Thus
$\Omega_{\mathrm{GW}}(k,x\rightarrow\infty)\propto a$ if we use Eq. \eqref{omegagw}. 
However, from Eq. \eqref{hsolution}, we see that the constant $6/5$ in Eq. \eqref{I_N}
contributes a constant to $h_{\bm k}$, so the contribution to
$h_{\bm k}'$ and the energy density is zero. 
This means that we should use the definition \eqref{rhogw1} to calculate $\Omega_{\text{GW}}$, otherwise the constant $6/5$ will be mistakenly accounted for if we use Eq. \eqref{omegagw}.
Therefore, the constant $6/5$ in Eq. \eqref{I_N} does not contribute to the energy density $\Omega_{\mathrm{GW}}$.
In other words, the constant in Eq. \eqref{I_N} does not represent a wave
and GWs come from those terms oscillating as $\sin x$ and $\cos x$.
After dropping the constant $6/5$, 
we have $I_{\mathrm{N}}(x\rightarrow \infty)\propto \cos x/x^2=\cos x/a$ leading to $\Omega_{\text{GW}}\propto a^{-1}$ and $\rho_{\text{GW}}\propto a^{-4}$, 
which behaves, as expected, as radiation in MD era. 
In \cite{Domenech:2020xin}, the authors obtained the above result by gauging away the constant term.
In summary, only the terms oscillating as $\sin x$ and $\cos x$ account for SIGWs.

\subsection{The gauge transformation}

Now we discuss the gauge transformation and how 
the energy density in other gauges can be derived from the result in the Newtonian gauge \cite{Lu:2020diy}.
We start from the infinitesimal coordinate transformation
$x^\mu\to x^\mu+\epsilon^\mu$ with $\epsilon^\mu=(\alpha,\delta^{ij}\partial_j\beta)$. 
For the discussion of SIGWs, we do not consider the vector degrees of freedom for the coordinate transformation, and the scalars $\alpha$ and $\beta$ are of first order.
Since the gauge transformation of tensor modes does not depend on the coordinate transformation of the same order, we do not need to consider the second-order gauge transformation.
For the second-order tensor perturbation, we have \cite{Malik:2008im, Bruni:1996im,Lu:2020diy}
 \begin{equation}
 \label{gaugetranf1}
     h_{ij}^{\mathrm{TT}}\to h_{ij}^{\mathrm{TT}}+\chi_{ij}^{\mathrm{TT}},
 \end{equation}
where
\begin{equation}
\label{xijtt}
\chi_{ij}^{\mathrm{TT}}(\bm x,\eta)=\int\frac{\mathrm{d}^3 k}{(2\pi)^{3/2}}e^{i\bm k\cdot \bm x}\left(\chi^+_{\bm k}(\eta) \mathbf e_{ij}^{+} +{\chi}^\times_{\bm k}(\eta)\mathbf e_{ij}^\times \right),
\end{equation}
\begin{equation}
    \label{chi_fourier}
\begin{split}
\chi^t_{\bm k}(\eta)=\frac{4}{k^2}\int\frac{\mathrm{d}^3p}{(2\pi)^{3/2}}\mathbf e^t_{ij}p^i p^j \zeta(\bm p)\zeta(\bm k-\bm p)I_\chi(u,v,x), 
\end{split}
\end{equation}
and
\begin{widetext}
\begin{equation}
 \label{Ichi}
 \begin{split}
   I_\chi(u,v,x)=\,\,&- \frac{9}{100uv}\left[\vphantom{\frac{u^2}{v^2}}2T_\alpha(ux)T_\sigma(vx)
   +2T_\alpha(vx)T_\sigma(ux)+2T_\alpha(ux)T_\alpha(vx)\right.\\
   &-4\left(\frac{u}{v}T_\psi(ux)T_\beta(vx)+\frac{v}{u}T_\psi(vx)T_\beta(ux)\right)\\
   &+\frac{1-u^2-v^2}{uv}\Bigl(T_\beta(ux)T_E(vx)+T_\beta(vx)T_E(ux)
   +T_\beta(ux)T_\beta(vx)\Bigr)\\
   &+\frac{8}{x}
   \left(\frac1v T_\alpha(ux)T_E(vx)+\frac1u T_E(ux)T_\alpha(vx)\right.\left.\left.+\frac1v T_\alpha(ux)T_\beta(vx)+\frac1u T_\beta(ux)T_\alpha(vx)\right)\right].
 \end{split}
 \end{equation}
\end{widetext}
We have symmetrized $I_\chi(u,v,x)$ under the interchange $u\leftrightarrow v$.
The transfer functions $T_\alpha$ and $T_\beta$ for the scalar parts $\alpha$ and $\beta$ of the infinitesimal coordinate transformation $\epsilon^\mu$ to the  first order  are
 \begin{gather}
 \label{defs.}
      \alpha(\bm k,x)=\frac{3}{5}\zeta(\bm k)\frac{1}{k}T_\alpha(x),\\
   \beta(\bm k,x)=\frac{3}{5}\zeta(\bm k)\frac{1}{k^2}T_\beta(x).
 \end{gather}
From the gauge conditions, it is not hard to find out the coordinate transformation between two gauges and the solutions for $\alpha$ and $\beta$.
With the gauge transformation \eqref{gaugetranf1} and the result for SIGWs in the Newtonian gauge, it is straightforward to derive the semianalytic expression for SIGWs in other gauges without performing the detailed calculation in that gauge.
In particular, combining Eqs. \eqref{hijkeq1}, \eqref{hsolution}, \eqref{gaugetranf1}, \eqref{xijtt}, and \eqref{chi_fourier}, we get the following gauge transformation \cite{Lu:2020diy}
\begin{equation}\label{hchi}
\begin{split}
&h^t_{\bm k}\rightarrow h^t_{\bm k}+\chi^t_{\bm k}\\&
=\frac{4}{k^2}\int\frac{\mathrm{d}^3p}{(2\pi)^{3/2}}\mathbf e^t_{ij}(\bm k)p^i p^j\zeta(\bm p)\zeta(\bm k-\bm p)\\&
\qquad\qquad \times\Bigl(I(u,v,x)+I_\chi(u,v,x)\Bigr).
\end{split}
\end{equation}
This powerful transformation rule allows us to quickly transform the solution of second-order tensor perturbation 
in one gauge to other gauges with the replacement of $I(u,v,x)$ by $I(u,v,x)+I_{\chi}(u,v,x)$. 
By setting the initial gauge to be the Newtonian gauge, 
we obtain $I_{\chi}(u,v,x)$ in other gauges as
\begin{align}\label{IN}
  I_\mathrm{N}(u,v,x) \to I_{\mathrm N}(u,v,x)+I_\chi(u,v,x),
\end{align}
where
\begin{equation}
\label{ichi_newton}
\begin{split}
   I_\chi(u,v,x)= &- \frac{9}{100uv}\left[\vphantom{\frac{u^2}{v^2}}
   2T_\alpha(ux)T_\alpha(vx) \right.\\
   &-4\left(\frac{u}{v}T_\mathrm{N}(ux)T_\beta(vx)
   +\frac{v}{u}T_\mathrm{N}(vx)T_\beta(ux)\right)\\
   &+\frac{8}{x}
   \left(\frac1v T_\alpha(ux)T_\beta(vx)+\frac1u T_\beta(ux)T_\alpha(vx)\right)\\
&   \left.+\frac{1-u^2-v^2}{uv}T_\beta(ux)T_\beta(vx) \right].
\end{split}
\end{equation}
$I_\chi$ can be obtained by substituting the transfer functions $T_\sigma=T_E=0$ and $T_\psi=T_\text{N}$
in the Newtonian gauge into Eq. \eqref{Ichi}.
The coordinate transformations from the Newtonian
gauge to the other gauges give the transfer functions $T_\alpha$ and $T_\beta$.

\section{The kernel in different gauges}
\label{sec.3}

The goal of this section is to compute the analytic expressions for the kernel $I_{\chi}$  in  six different gauges from the Newtonian gauge
 by using Eq. \eqref{IN}.
 
\subsection{Synchronous gauge}
In this subsection, we first calculate the kernel directly by using Eq. \eqref{I_int} and then confirm that it is the same as that obtained from the gauge transformation with Eq. \eqref{IN}.  

In  the synchronous gauge, $\phi=B=0$. 
The equation for the transfer function $T_E$ in the synchronous gauge is
\begin{equation}
\label{sync:teeq}
x^3T_{E}^{****}+8x^2T_E^{***}+8x T_E^{**}-8T_E^*=0.
\end{equation}
The general solution for the transfer function $T_E$ is
\begin{equation}
\label{sync:tesol}
T_E(x)=\mathcal{C}_1+\frac{x^2}{2}{\mathcal C_2}-\frac{\mathcal C_3}{x}-\frac{\mathcal C_4}{3x^3},
\end{equation}
where $\mathcal{C}_i$ are integration constants.
Note that there are two gauge modes in Eq. \eqref{sync:tesol}
because of the residual gauge freedom in the synchronous gauge \cite{Wang:2018yql,Press:1980is,Bucher:1999re,Bednarz:1984dn,Ma:1995ey}.
To identify these two gauge modes, we take
the residual gauge transformation \cite{Press:1980is,Bucher:1999re}
\begin{equation}
\label{sync:transf}
\begin{split}
\alpha&=\frac{{\mathcal{C}_5}}{x^2},\\
\beta&=-\frac{\mathcal{C}_5}{x}+\mathcal{C}_6,
\end{split}
\end{equation}
and after that we are still in synchronous gauge where $\phi=B=0$.
From the transformation \eqref{gauge:etransf}, we see that the constant
$\mathcal{C}_6$ term in $\beta$ contributes
to the integration constant $\mathcal{C}_1$ in Eq. \eqref{sync:tesol} and the $\mathcal{C}_5$ term in
$\beta$ contributes to the $1/x$ term in Eq. \eqref{sync:tesol}.
Therefore, $\mathcal{C}_1$ and $\mathcal{C}_3$ terms in Eq. \eqref{sync:tesol} are just pure gauge modes and we can eliminate them by substituting  $\mathcal{C}_1=\mathcal{C}_3=0$.
Now we determine the remaining integration constants from the initial condition.
At the initial time $x=0$, assume that $T_E(0)$ is finite, 
we get $\mathcal{C}_4=0$.
Then we use the initial condition of the gauge-invariant Bardeen potential to fix the constant $\mathcal{C}_2$.
The gauge-invariant Bardeen potential in synchronous gauge is
\begin{equation}
\Phi=-\mathcal{H}E'-E'',
\end{equation}
so the transfer function $T_\Phi$ is
$T_\Phi= -3{\mathcal{C}_2}$.
From the initial condition $T_\Phi(0)=T_{\mathrm{N}}=1$, we derive $\mathcal{C}_2=-1/3$.
After eliminating the gauge modes, we get the transfer function $T_E$ as \cite{Inomata:2019yww}
\begin{equation}
\label{synte_psi}
T_E(x)=-\frac{x^2}{6}.
\end{equation}
The transfer function $T_\psi$ is \cite{Ma:1995ey,Inomata:2019yww}
\begin{equation}
\label{syn_tpsi}
T_\psi(x)=-T_E^{**}(x)-\frac{4}{x}T_E^*(x)=\frac{5}{3}.
\end{equation}
Substituting Eqs. \eqref{syn_tpsi} and \eqref{synte_psi} into Eq. \eqref{I_int}, we get
\begin{equation}
\label{Isyn}
\begin{split}
        I_{\mathrm{syn}}(u,v,x)=& \frac{x^2}{400}\Bigl(
-88+(-1+u^2+v^2)x^2\Bigr) \\&+ \frac{6 \left(x^3-3 \sin x+3 x \cos x\right)}{5 x^3}.
\end{split}
\end{equation}

Next we derive the kernel by using the gauge transformation.
The coordinate transformation from the Newtonian gauge to the synchronous gauge is 
\begin{equation}
\label{synalpha}
\begin{split}
  \alpha(\bm k,x)&=\frac35\zeta(\bm k)\frac1kT_\alpha(x),\\
  \beta(\bm k,x)&=\frac3 5\zeta(\bm k)\frac{1}{k^2}T_\beta(x),
\end{split}
\end{equation}
where 
\begin{equation}
\label{T_beta_syn}
\begin{split}
  T_\alpha(x)=-  \frac{x}{3},\\
  T_\beta(x)=- \frac{x^2}{6} .
\end{split}
\end{equation}
Substituting Eq. \eqref{T_beta_syn} into Eq. \eqref{ichi_newton}, we get
\begin{equation}
\label{synkernel}
I_{\chi}(u,v,x)
=\frac{x^2}{400}\Bigl[
-88+(-1+u^2+v^2)x^2\Bigr].
\end{equation}
We confirm that $I_{\mathrm{syn}}=I_{\text{N}}+I_\chi$. In the next subsections, we derive the kernels by using the gauge transformation \eqref{IN} only. 

It is noteworthy that in Eq. \eqref{synkernel}, there is no oscillating term to represent GWs. Therefore,
the kernel $I_{\chi}(u,v,x)$ does not contribute to SIGWs
and $\Omega_{\text{GW}}$ in both the Newtonian gauge and the synchronous gauges
are the same.

\subsection{Comoving orthogonal gauge}

In the comoving gauge, $\delta V=0$ and $B=0$ \cite{Malik:2008im}.
This gauge also retains a residual coordinate transformation $\beta=\mathcal{C}$
which corresponds to arbitrary choice of the origin of the spatial coordinates.
For the time coordinate transformation, the variable $\alpha$ is given by
\begin{align}
\label{alphacomov}
  \alpha=\frac{\mathcal{H}\phi_{\mathrm N}+\phi_{\mathrm N}^\prime}{\mathcal{H}'-\mathcal{H}^2}.
  \end{align}
From the above expression, we get the transfer function
\begin{equation}
\label{transacomov}
  T_\alpha(x)=-x/3.
\end{equation}
The general solution of  the transfer function $T_\beta$ is
\begin{equation}
\label{cotransfer}
  T_\beta(x) =-
\frac{x^2}{6}+\mathcal{C}.
\end{equation}
The last constant $\mathcal{C}$ term is a pure gauge mode, we can choose $\mathcal{C}=0$ so that ${T_{\beta}(x=0)=0}$.
Combining Eqs. \eqref{ichi_newton}, \eqref{transacomov} and \eqref{cotransfer}, we get
\begin{equation}
I_{\chi}\left(u,v,x\right)=\frac{x^2}{400}\Bigl[
-88+(-1+u^2+v^2)x^2\Bigr].
\end{equation}
It is interesting to note that the kernel in the comoving orthogonal gauge is identical with that in synchronous gauge and there is no oscillation, 
so this kernel does not contribute to SIGWs. 
For SIGWs, $\Omega_{\text{GW}}$ in the comoving
orthogonal gauge is the same as those in the Newtonian and synchronous gauges.

\subsection{Uniform curvature gauge}

The uniform curvature gauge demands $\psi=E=0$.
The coordinate transformation
from the Newtonian gauge to the uniform curvature gauge is
\begin{align}
\label{alphaunif}
  \alpha&=\frac{\phi_{\mathrm{N}}}{\mathcal{H}}=\frac35\zeta(\bm k)\frac1k T_\alpha(x),\\
  \beta&=0,
\end{align}
where the transfer function  $T_\alpha$ is
\begin{equation}
  T_\alpha= xT_\text{N}/2.
\end{equation} 
Substituting these results of the transfer functions  into  Eq. \eqref{ichi_newton}, 
we get
\begin{equation}
\label{Iuc}
    I_\chi(u,v,x)=-\frac{9x^2}{200}.
\end{equation}
This term is a growing mode and there is no oscillation, so it does not contribute to SIGWs.
For SIGWs, we find that $\Omega_{\text{GW}}$ in the uniform curvature
gauge is the same as that in the Newtonian gauge.

\subsection{Comoving gauge (total matter gauge)}

The comoving gauge (also referred as the total matter gauge \cite{Malik:2008im}) is defined by the condition,
$\delta V=E=0$.
The transfer functions for the coordinate transformation from the Newtonian gauge to the comoving gauge are
\begin{equation}
\label{transferTM}
  T_\alpha(x)=-\frac{x}{3},
\end{equation}
and $T_\beta(x)=0$.
Substituting  Eq. \eqref{transferTM} into  Eq. \eqref{ichi_newton}, we get
\begin{align}
\label{Ichi_tm}
I_\chi(u,v,x)=-\frac{x^2}{50}.
\end{align}
This term is a growing mode and there is no oscillation, so it does not contribute to SIGWs.
For SIGWs, we obtain the same $\Omega_{\text{GW}}$ in the total matter gauge as that in the Newtonian gauge.

\subsection{Uniform density gauge}

The uniform density gauge is defined by the condition, $\delta\rho=E=0$.
The transfer functions for the coordinate transformation
from the Newtonian gauge to this gauge are
\begin{equation}
\label{transferfuncUD}
  T_\alpha(x)=-\frac{12x +x^3}{36},
\end{equation}
and $T_\beta(x)=0$.
Substituting  Eq. \eqref{transferfuncUD} into  Eq. \eqref{ichi_newton}, we get
\begin{equation}
\label{IalphaUD}
  I_{\chi}(u,v,x)=-\frac{x^2 (12+u^2 x^2) (12+v^2x^2)}{7200}.
\end{equation}
This term is a growing mode without oscillation, so it does not contribute to SIGWs.
For SIGWs, we find that $\Omega_{\text{GW}}$ in the uniform density
gauge is the same as that in the Newtonian gauge.

\subsection{Uniform expansion gauge}

At last, we consider the uniform expansion gauge,
$3(\mathcal{H}\phi+\psi^\prime)+k^2\sigma=0$ and $E=0$ \cite{Hwang:2017oxa}. 
The transfer functions for the coordinate transformation
from the Newtonian gauge to this gauge are
\begin{equation}
\label{transferfuncuniexp}
  T_\alpha(x)=-\frac{6x}{18+x^2},
\end{equation}
and $T_\beta(x)=0$.
Substituting  Eq. \eqref{transferfuncuniexp} into  Eq. \eqref{ichi_newton}, we get
\begin{equation}
\label{IalphaUniexp}
  I_{\chi}(u,v,x)=-\frac{162 x^2}{25(u^2 x^2+18) (v^2 x^2+18)}.
\end{equation}
At late times, this kernel is a decaying mode without oscillation, 
so it does not contribute to SIGWs.
For SIGWs, we obtain the same $\Omega_{\text{GW}}$ in the uniform expansion
gauge as that in the Newtonian gauge.

\subsection{Monochromatic power spectrum}

For the monochromatic power spectrum, $\mathcal{P}_\zeta=A_\zeta\delta(\ln k-\ln k_*)$,
the energy density for free SIGWs in MD is
\begin{equation}
\label{omega_md}
\begin{split}
    \Omega_{\mathrm{GW}}(k,x)&=\frac{1}{48x^2}A_\zeta^2
    \left(
    1-\left(\frac{k}{2k_*}\right)^2
    \right)^2\Theta(2k_*-k)\\
    &\propto 1/a,
\end{split}
\end{equation}
where $\Theta(x)$ is the Heaviside theta function,  and $k_*$ is the wave-number of the peak in the power spectrum.
Since the energy density of matter decays as $a^{-3}$ and the
energy density of SIGWs decays as $a^{-4}$, 
so $\Omega_{\text{GW}}$ decays as $1/a$ in MD.

\section{Conclusion}
\label{sec4}
In this paper, we study the solutions to scalar-induced tensor perturbations in various gauges during MD. 
Since the time dependence of SIGWs $h^t_{\bm k}(\eta)$ lies
in the integral kernel $I(u,v,x)$, 
we explicitly calculate the analytical expressions for the kernels $I_{\text{N}}$
in the Newtonian and $I_{\mathrm{syn}}$ in the synchronous gauges. We also derive the relation between the kernels in different gauges
under the coordinate transformation and use the results to obtain
the analytical expressions for the kernels in six other gauges, 
namely the synchronous gauge,
the comoving orthogonal gauge, the uniform curvature gauge, 
the total matter gauge, the uniform density gauge
and the uniform expansion gauge. 
The direct calculation of $I_{\mathrm{syn}}$ in the synchronous gauge 
confirms that it is the same as that obtained from $I_{\text{N}}$ 
using the gauge transformation.
There are two residual gauge modes in the synchronous gauge
and one residual gauge mode in the comoving orthogonal gauge.
After identifying and eliminating the gauge modes, 
we find that the kernels in the synchronous gauge and the comoving orthogonal gauge are the same.

Although the derived kernels are different in different gauges, 
the difference is from either growing or decaying modes with the behavior $x^n$ or $a^{n/2}$.
For SIGWs with the oscillating behaviors $\sin x$ and $\cos x$, 
we find that at late times $\rho_{\text{GW}}\propto a^{-4}$, and $\Omega_{\text{GW}}\propto a^{-1}$
in all seven gauges, i.e., SIGWs behave the same as radiation in MD.
With the analytical expression for the kernel, we give the analytical result
of the energy density $\Omega_{\text{GW}}$ for the monochromatic power spectrum.
In conclusion, SIGWs are gauge independent, and it is convenient to calculate $\Omega_{\text{GW}}$ in the Newtonian gauge in practice.
Our results are helpful for the probe of the thermal history of the universe with SIGWs. 

\begin{acknowledgments}
This work is supported by the National Natural Science Foundation of
China under Grant No. 11875136, the Major Program of the National Natural Science
Foundation of China under Grant No. 11690021 and the National Key Research and Development Program of China under Grant No. 2020YFC2201504.
\end{acknowledgments}
\appendix

\section{GAUGE TRANSFORMATION}
\label{app.A}

\subsection{The energy-momentum tensor}
A perfect fluid has the stress energy-momentum tensor of the form
\begin{align}
\label{stresstensor}
  T_{\mu\nu}=(\rho+P)U_\mu U_\nu+Pg_{\mu\nu}+\Pi_{\mu\nu},
\end{align}
where the anisotropic stress  ${\Pi}_{\mu\nu}$ is assumed to be zero.
The first-order perturbations of the velocity $U_\mu$,
the energy density, the pressure,
and the anisotropic stress are $\delta U_\mu$, $\delta\rho$, $\delta P$, and $\delta\Pi_{ij}$, respectively. The first-order velocity perturbation
$\delta U_\mu$ is decomposed via $\delta U_\mu=a[\delta V_0,\delta V_{,i}+\delta V_i]$ with $\delta V_{i,i}=0$. Notice $\delta V$ we defined here relates to the perturbations $v_{\mathrm M}$ and $B_{\mathrm M}$ in Ref. \cite{Malik:2008im} by $\delta V=v_{\mathrm M}+B_{\mathrm M}$.

\subsection{Gauge transformations}
\label{app.B}
One may perform a gauge transformation of the form 
$x^\mu\rightarrow \tilde{x}^\mu=x^\mu+\epsilon^\mu(x)$ 
under the general infinitesimal coordinate transformation, with
$\epsilon^\mu=[\alpha,\delta^{ij}\partial_j\beta]$
. The first-order gauge transformations are written as
\begin{align}
\tilde{\phi}=&\phi + \mathcal{H}\alpha + \alpha^\prime,\\
\tilde{\psi}=&\psi -\mathcal{H}\alpha,\\
\tilde{\sigma}=&\sigma+\alpha,\\
\tilde{B}=&B -\alpha +\beta',\\
\label{gauge:etransf}
\tilde{E}=&E+\beta,\\
\delta\tilde{\rho}=&\delta\rho+{\rho}_0^\prime\alpha,\\
\delta\tilde{P}=&\delta P+P_0^\prime\alpha,\\
\delta\tilde{V}=&\delta V-\alpha,\\
\delta\tilde{\Pi}=&\delta\Pi,
\end{align}
where $\Pi$ is the scalar part of the (trace-free) anisotropic stress.
One can obtain two  gauge-invariant Bardeen potentials by using the above gauge transformation \cite{Bardeen:1980kt}
\begin{align}
\label{bardeenphi}
\Phi=&\phi-\mathcal{H}\sigma-\sigma^\prime,\\
\label{bardeenpsi}
\Psi=& \psi+\mathcal{H}\sigma.
\end{align}
The gauge transformation for the SIGWs under the infinitesimal coordinate transformation is $h_{ij}^{\mathrm{TT}}\to h_{ij}^{\mathrm{TT}}+\chi^{\mathrm{TT}}_{ij}$,
where
\begin{equation}
\label{hgaugetrans}
\begin{split}
\chi_{ij}=&2\left[\left(\mathcal{H}^2+\frac{a^{''}}{a}\right)\alpha^{2}
+\mathcal{H}\left(\alpha\alpha'+\alpha_{,k}\epsilon^{k}\right)\right]\delta_{ij}\\
&+4\left[\alpha\left(C'_{ij}+2\mathcal{H}C_{ij}\right)+C_{ij,k}\epsilon^k+
C_{ik}\epsilon^k_{,j}+C_{jk}\epsilon^k_{,i}\right]\\
&+2\left(B_{i}\alpha_{,j}+B_{j}\alpha_{,i}\right)
+4\mathcal{H}\alpha\left(\epsilon_{i,j}+\epsilon_{j,i}\right)
-2\alpha_{,i}\alpha_{,j}\\&+\left(\epsilon_{i,jk}+\epsilon_{j,ik}\right)\epsilon^k +\epsilon_{i,k}\epsilon^{k}_{,j}+\epsilon_{j,k}\epsilon^{k}_{,i}+\epsilon'_{i}\alpha_{,j}+\epsilon'_{j}\alpha_{,i},\\&
+2\epsilon_{k,i}\epsilon^{k}_{,j}+
\alpha\left(\epsilon'_{i,j}+\epsilon'_{j,i}\right)
\end{split}
\end{equation}
and  $C_{ij}=-\psi\delta_{ij}+E_{,ij}$.


%

\end{document}